\pdfoutput=1
\documentclass[final]{siamltex}
\usepackage{graphicx}
\usepackage{algorithm, algorithmic}
\usepackage{amsmath}
\usepackage{amsfonts}
\usepackage{setspace}
\usepackage{listings}
\usepackage{courier}

\usepackage[usenames,dvipsnames]{color} 

\def\eg{e.\,g.}

\def\ie{i.\,e.}

\title{Optimizing CUDA Code By Kernel Fusion---Application on BLAS}

\author{Ji\v{r}\'{\i} Filipovi\v{c}\footnotemark[2]
\and Mat\'{u}\v{s} Madzin\footnotemark[3]
\and Jan Fousek\footnotemark[3]
\and Lud\v{e}k Matyska\footnotemark[2]}

\begin{document}

\maketitle

\footnotemark[2]{Institute of Computer Science, Masaryk University, Botanick\'{a} 68a, 602 00 Brno, Czech Republic}
\footnotemark[3]{Faculty of Informatics, Masaryk University, Botanick\'{a} 68a, 602 00 Brno, Czech Republic}

\begin{abstract}

Modern GPUs are able to perform significantly more arithmetic operations than transfers of a single word to or from global memory. Hence, many GPU kernels are limited by memory bandwidth and cannot exploit the arithmetic power of GPUs. However, the memory locality can be often improved by kernel fusion when a sequence of kernels is executed and some kernels in this sequence share data.

In this paper, we show how kernels performing map, reduce or their nested combinations can be fused automatically by our source-to-source compiler. To demonstrate the usability of the compiler, we have implemented several BLAS-1 and BLAS-2 routines and show how the performance of their sequences can be improved by fusions. 
Compared to similar sequences using CUBLAS, our compiler is able to generate code that is up to $2.61\times$ faster for the examples tested. 

\end{abstract}

\begin{keywords}
GPU, CUDA, BLAS, Kernel fusion, Code generation, Automated tuning
\end{keywords}

\begin{AMS}
68W10, 68N19, 15A99
\end{AMS}

\section{Introduction and Motivation}

Today's accelerators, such as CUDA GPUs, are able to perform tens of arithmetic operations in the time that it takes for a word to be read from or written to global memory. Moreover, the dominance of arithmetic power over memory bandwidth grows with each new hardware generation\footnote{The first CUDA processor, G80, has flop-to-word ratio about 24, GT200 has 27, GF110 has 33 and GK110 has 63.}. The input and output of each GPU kernel (\ie{} the subprogram executed on GPU) has to be stored in the global memory. Thus, many kernels  with low flop-to-word ratio are memory-bound. 
When such kernels executed in sequence share some data, performance may be improved by placing the shared data in some significantly faster on-chip memory. Although global memory is cached in new GPUs, caches usually cannot hold whole output of the kernel. However, the memory locality can be improved by fusing these kernels into a larger ones and placing shared data into on-chip memory explicitly.

The number of possible fusions is high as each fusion is created according to sequence of kernel calls and data dependency between them. Thus,  re-usability of fused kernels is limited.
Because of this, it is \textit{impractical to produce libraries} consisting of already-fused kernels. Instead, it is more practical to use the library of simple and re-usable kernels and \textit{automatically generate fusions} when the sequence of kernel calls is given. 

It is difficult to fuse generic kernels automatically, but automation of fusion becomes possible when the type of operations performed by kernels is limited. In this paper, we study automatic fusions of kernels performing \textit{map}, \textit{reduce} or their nested combination. In our approach, the function applied by map or reduce can run in multiple threads. Thus, it can efficiently process larger amount of data, which allows common optimization of memory locality, such as tiling.

In this paper, we present kernel fusion as an optimization method and show how it can be automated by our source-to-source compiler when the type of fused kernels is restricted to map and reduce. The compiler works with a~\textit{library of elementary functions} and a~\textit{script} calling functions from the library. It fuses selected functions to improve their performance and preserve the semantics defined by the script.  We note that fusing all kernels cannot always maximize performance. Thus, the compiler searches and prunes the optimization space to find efficient fusions.

We address two main use cases by our approach.
\begin{itemize}
        \item \textit{Using fusion-equipped libraries}. Some general purpose library can be implemented to be usable with our compiler. In that case, library users can write only script calling library functions without the need to care about their GPU-specific implementation. The advantage of this approach is that library functions are automatically fused by our compiler, improving their performance. 
        \item \textit{Simplification of fusion optimization}. In some cases, it is meaningful to develop both the script and the library (even if is not widely reusable) and use our compiler to find efficient fusions. First, many combinations of library function calls may be needed which makes the manual fusion time demanding and error-prone. Second, the optimization of code-to-kernels distribution may be hard (one such example is presented in our previous paper~\cite{filipovic2012automatically}).
\end{itemize}

To demonstrate the performance benefit of kernel fusions generated by our compiler, \textit{we have accelerated several sequences of BLAS} (Basic Linear Algebra Subprograms) routine calls. BLAS is a library of linear algebra routines, which is frequently used in scientific computation and is believed to be well-optimized. The BLAS-1 (vector-vector) and BLAS-2 (matrix-vector) routines are memory-bound, thus their sequences are good candidates to be improved by fusions \cite{belter2009automating, howell2008cache}. We show that fusing several BLAS routines into a single kernel can significantly improve performance by reducing the number of memory transfers. This performance improvement cannot be achieved by tuning unfused kernels separately. 


The rest of the paper is structured as follows. The overview of work related to our research is given in Section \ref{sect:related}. 
The general discussion about performance impact of kernel fusion as well as its automation can be found in Section \ref{sect:kerfus}, whereas Section \ref{sect:compilation} describes the compiler allowing automatic fusions. The performance of a code generated by our compiler is evaluated in Section \ref{sect:eval}. The Section \ref{sect:conclusion} concludes the paper and sketch the future work.

\section{Related Work}
\label{sect:related}


The code-to-kernel distribution can be optimized by kernel fusion, or by generation of kernels of optimized size from a code which is not explicitly divided into kernels (monolithic implementation or some high-level language).

The kernel fusion is allowed in some frameworks working with algorithmic skeletons. Algorithmic skeletons that allow automatic parallelization are predefined higher-order functions performing given user-defined first-order functions~\cite{cole1989algorithmic, velez2010survey}. The SkeTo framework automatically fuse skeletons to spare global memory transfers~\cite{sato2009skeletal}. The fusion is possible also in Thrust~\cite{hoberock2009thrust}, but the programmer has to explicitly set the kernels to be fused. The significant difference of our approach is that first-order functions can be parallel, which allows them to process larger data (\eg{} small tensors~\cite{filipovic2012automatically} or matrix tiles), whereas user-defined functions executed by skeletons are serial. Second difference is that we search fusion optimization space to discard suboptimal fusions.

In array programming, one defines the transformations of whole arrays using element-wise operations, reduction etc.~\cite{iverson1962programming} Although array and skeletal programming introduce different programming models, the transformations of arrays performs usually similar operations as skeletons and there is similar opportunity to perform several transformations within single kernel. The Barracuda compiler is able to fuse arrays and perform operations on these arrays in a~single kernel~\cite{larsen2011simple}. The fusions are performed whenever it is possible, without considering on-chip resources consumption. A similar fusion mechanism is implemented in Copperhead~\cite{catanzaro2011copperhead}, which is a high-level data-parallel language embedded in Python. It seems that both Barracuda and Copperhead do not discard suboptimal fusions. A programmer cannot write the native code of the transformation applied to array's elements (\ie{} first-order functions), thus he or she cannot explicitly define parallel per-element code or implement any low-level optimization, which is possible in our approach. On the other hand, our approach is more low-level, as our compiler fuses functions written in C for CUDA.

A tool by Gulati and Khatri~\cite{gulati2009automated} automatizes the partitioning of the input code into kernels and automatically generates the code of output kernels. The input code performs serial computation, the output code performs the computation multiple times in parallel (thus it is application of map function). The paper shows that the optimization of resource usage by partitioning of the code into several kernels may in some cases improve the performance over monolithic implementation even if the data locality is worse in the partitioned code, which agree with our results.

A fusion method improving energy efficiency of CUDA kernels is proposed in~\cite{wang2010kernel}. This method does not aim at improving the execution times of kernels, as kernels are fused without improving data locality. Instead, it aims at balance the demand for hardware resources, resulting in lower power consumption, but similar execution times. Similarly to our method, the previously-implemented kernels are fused instead of automatic parallelization.

The idea of optimizing sequences of BLAS functions by their fusion is not new, however, to the best of author's knowledge, no system allowing fusions on GPUs has been published. Belter at al.~\cite{belter2009automating} introduce a~BTO BLAS compiler, which is able to fuse BLAS functions targeting modern CPUs. The DESOLA active library, presented by Russell at al.~\cite{russell2011desola}, performs fusions  in time the BLAS functions are called, \ie{} without a~previous compilation. 
The main difference between our research and those presented in~\cite{belter2009automating} and~\cite{russell2011desola} is that we target GPU architecture, thus the technique of the fusion significantly differs. We fuse parallel kernels instead of loops, which requires different techniques to perform the fusion correctly.
Moreover, the optimization space search and performance prediction also changes due to different nature of GPUs.
Our approach addresses multiple types of computational problems, whereas BTO BLAS and DESOLA focus only on BLAS. Our approach uses the hand-tuned routines, whereas BTO BLAS uses high-level description of BLAS routines and DESOLA implements initial BLAS functions (which are further optimized automatically) in language similar to C, without any architecture-specific optimizations.

In our previous papers, we have introduced basic principles of our compiler~\cite{fousek2011automatic} and its non-trivial application together with improved efficiency of data exchanging between fused kernels~\cite{filipovic2012automatically}. Nevertheless, the compiler introduced in these papers was restricted on map kernels, which significantly limits its applicability. In this paper, we discuss fusion of nested map and reduce and show the structure of generated code and process of its generation in deeper detail.

\section{Kernel Fusion}
\label{sect:kerfus}

In this section, we discuss kernel fusion in more detail, but still as a  general concept, \ie{} independently of the design and implementation decisions made for our compiler. First, the performance advantages and disadvantages of fusions are discussed. Second, the properties of map and reduce functions allowing their automatic fusion are described. Finally, the implementation of BLAS routines as fusible kernels is introduced.

\subsection{Fusion as an Optimization Method}

The main advantage of fusion is the improvement of memory locality. In CUDA, each kernel has to read its input from and store its output to off-chip global memory. When two kernels share some data, their fusion can hold shared data in on-chip memory---registers or shared memory. 

\begin{figure}[t]
\centering
\includegraphics[width=.4\hsize]{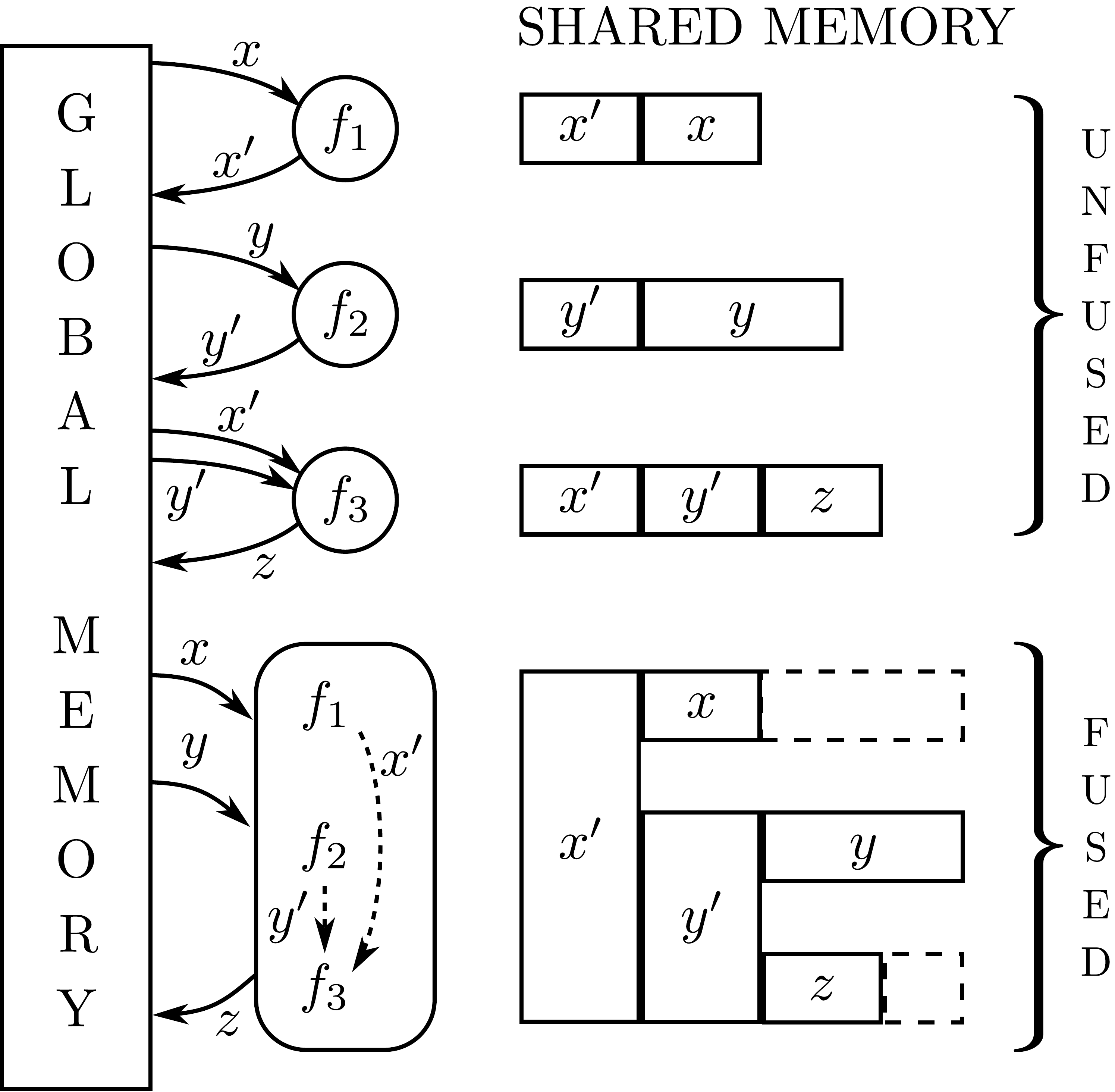}
\caption{Computation of $z = f_3(f_1(x), f_2(y))$ as $x'=f_1(x)$, $y'=f_2(y)$, $z = f_3(x',y')$. Left: data movement of unfused and fused versions. Right: On-chip memory allocation in unfused and fused versions.}
\label{fig:fusion-ilustration}
\end{figure}

Consider the example depicted in Figure~\ref{fig:fusion-ilustration} left, where $z = f_3(f_1(x), f_2(y))$ is evaluated. When $f_1, f_2$ and $f_3$ are fused into a single kernel, the results computed by $f_1$ and $f_2$ can be held in on-chip memory and immediately used by $f_3$. Otherwise, the outputs of $f_1$ and $f_2$ have to be stored in global memory and loaded by $f_3$. If the performance of $f_1$, $f_2$ or $f_3$ is bounded by global memory bandwidth, the fusion increases performance by reducing global memory data transfers.

An additional benefit of kernel fusion is the reduction of kernel launch overhead (a lower number of kernels are launched). Moreover, the fused kernels are more complex, thus the optimizing compiler has more room for optimizing the instructions, such as common subexpression elimination (\eg{} data indexing can be the same or similar for multiple functions included in single fusion), loop fusion, etc.

Besides the performance improvements mentioned above, fusion may also decrease performance. The occupancy of the GPU (the number of warps that can concurrently run on the GPU) must be sufficient to hide the memory latency \cite{nvidia2011nvidia}. When a kernel requires too much on-chip memory, occupancy is limited and the memory latency can decrease performance. When such a kernel is fused with another, occupancy is limited for the whole fused kernel. Thus, it is possible that the overall performance is better when the kernel which limits occupancy is not fused.

Another factor that can limit occupancy is the storage of additional intermediate data in on-chip memory. Consider the example mentioned above. In the fused kernel, $f_1$ and $f_2$ have to be performed before $f_3$ in any ordering. Suppose that $f_1$ is performed before $f_2$. In this case, when $f_2$ is performed, the result of $f_1$ must be held in on-chip memory, thus at least for $f_2$ the consumption of on-chip memory is higher compared to the unfused version. This example is depicted in Figure~\ref{fig:fusion-ilustration} right, where the execution of every unfused kernel would consume less on-chip memory compared to the fusion.

Finally, the optimal number of threads processing data elements can vary for different kernels. When such kernels are fused, some of them have to use a suboptimal number of threads, or some threads idle in part of the computation (but hold on-chip resources), thus fusion may decrease performance.

As we have shown, kernel fusion may increase as well as decrease the performance. The number of possible fusions and their combinations is large (see Table~\ref{tab:prediction} or \cite{filipovic2012automatically}) and a manual search for the best-performing one is time-consuming and error prone. Thus, the automatic generation of efficient fused code is necessary.

\subsection{Kernel Fusibility}
\label{subsect:kerfus}

To fuse two kernels, one has to correctly glue kernel codes into a~single kernel preserving the original functionality. The automatic fusion of generic kernels is difficult for two main reasons.
\begin{itemize}
        \item \textit{On-chip memory is distributed.} Some data, which was originally exchanged via global memory, resides in on-chip memory in fused kernel. This data can be transfered via on-chip memory when the following holds for all kernels to be fused: (i) kernels use the same mapping of threads to exchanged data placed in registers and (ii) kernels use the same mapping of thread blocks to exchanged data placed in shared memory. Thus, the automatic analysis of this mapping and its modification is needed.
        \item \textit{Global barrier is not available inside kernel}. Kernel execution creates a~global barrier, which cannot be generally implemented within a~kernel. Two or more kernels can be fused only if this global barrier is not necessary, \ie{} it can be replaced by a~local barrier or avoided entirely. Thus, it is needed to automatically determine whereas the global barrier can be avoided.
\end{itemize}

In our paper, we have restricted the types of kernels to map and reduce and their nested combinations (mapped map, or mapped reduce -- a map function cannot be used as a reduction operator). These kernels have a~wide range of applications as map and reduce have sufficient expressive power for many computing tasks. 
Also, map and reduce allow automatic fusion, as is shown below.
We note that the method of fusion is general, therefore more types of kernels could be fused automatically, although we currently do not support them. 

The idea of fusing GPU kernels performing map and reduce has been presented in several papers~\cite{sato2009skeletal, larsen2011simple, catanzaro2011copperhead, tarditi2006accelerator}. However, the map and reduce can execute parallel first-order functions in our case, which makes their fusions more complicated. Thus, we discuss in more details how to fuse them.

\subsubsection{Map Kernels}

Let $L_i = [e^i_1, e^i_2,\dotsc, e^i_m]$ is a~list of $m$ elements $e^i_1,\dotsc, e^i_m$. The map is a~higher-order function which applies a~given n-ary\footnote{Some languages uses map only for unary functions and introduce zipwith for n-ary functions.} function $f$ element-wise to all elements of the lists $L_1,\dotsc, L_n$, producing the list of results:

\begin{center}
        $map(f, L_1,\dotsc, L_n) = [f(e^1_1,\dotsc, e^n_1),\dotsc, f(e^1_m,\dotsc, e^n_m)]$
\end{center}

Suppose two data-dependent calls of map function $map(g, map(f, L))$, $L = [e_1, \dotsc, e_n]$. The mapped functions $f, g$ can be fused, \ie{} kernel performing $map(g \circ f, L)$ can be created, if and only if $\forall i \in [1, n]$, $f(e_i)$ and $g(f(e_i))$ run in the same (single) thread block. It guarantees that the result of $f$ can be transfered to $g$ via on-chip shared memory, as the shared memory is visible for all threads within the same block. It also guarantees that no global barrier is needed, as no data are exchanged between blocks.  We note that single instance of each mapped function has to fit into thread block, \ie{} has to use reasonable number of resources (threads, on-chip memory).

\subsubsection{Reduce Kernels}

Let~$\oplus$ be a~binary associative operator. The reduce is higher-order function applying given operator~$\oplus$ recursively to all elements of the list $L$ building a~resulting element.

\begin{center}
$reduce(\oplus, L) = e_1 \oplus e_2 \oplus e_3 \oplus \dotsb \oplus e_n$
\end{center}

The result of reduction is constructed using all elements $e_1 \dots e_n$. As multiple thread blocks are used to utilize GPU, a global barrier is needed to obtain the result of reduction. The important consequence of global barrier need is that the result of the reduction cannot be used in the same kernel where the reduction is computed. Nevertheless, we can fuse a reduction kernel with other kernel. Because of $\oplus$ associativity, a partial reduction can be computed locally per thread block without global barrier and thus reuse on-chip data (produced by map, or shared input of another reduce function). The final result of reduction is obtained after global barrier by reducing results of all partial reductions. 

We note that the final result of the reduction can be computed by several ways{ (i) by extra kernel, (ii) by the last running block of kernel performing partial reduction (global barrier is replaced by test of termination of all other blocks) or (iii) automatically after kernel is finished when atomic instructions are available.

\subsubsection{Local Barriers and Registers}

Let $f$ and $g$ are functions being fused. The thread-to-data mapping of $f, g$ is same if and only if each word transfered from $f$ to $g$ is stored in $f$ and loaded in $g$ by the same thread. As our mapped functions or reduce operators can be parallel, the thread-to-data mapping can differ in kernels being fused. In that case, data has to be transfered via shared memory and local barrier needs to be performed between kernel codes. 

The local barrier is not needed between $f$ and $g$ when the thread-to-data mapping is same. When all functions accessing data element $e$ access them with same thread-to-data mapping, and the access is not data-dependent\footnote{Data element can be placed in registers only if their indexing can be determined in compile time~\cite{nvidia2011nvidia}.}, the element $e$ can be stored in registers.

\begin{figure}[h]
\centering
\includegraphics[width=.5\hsize]{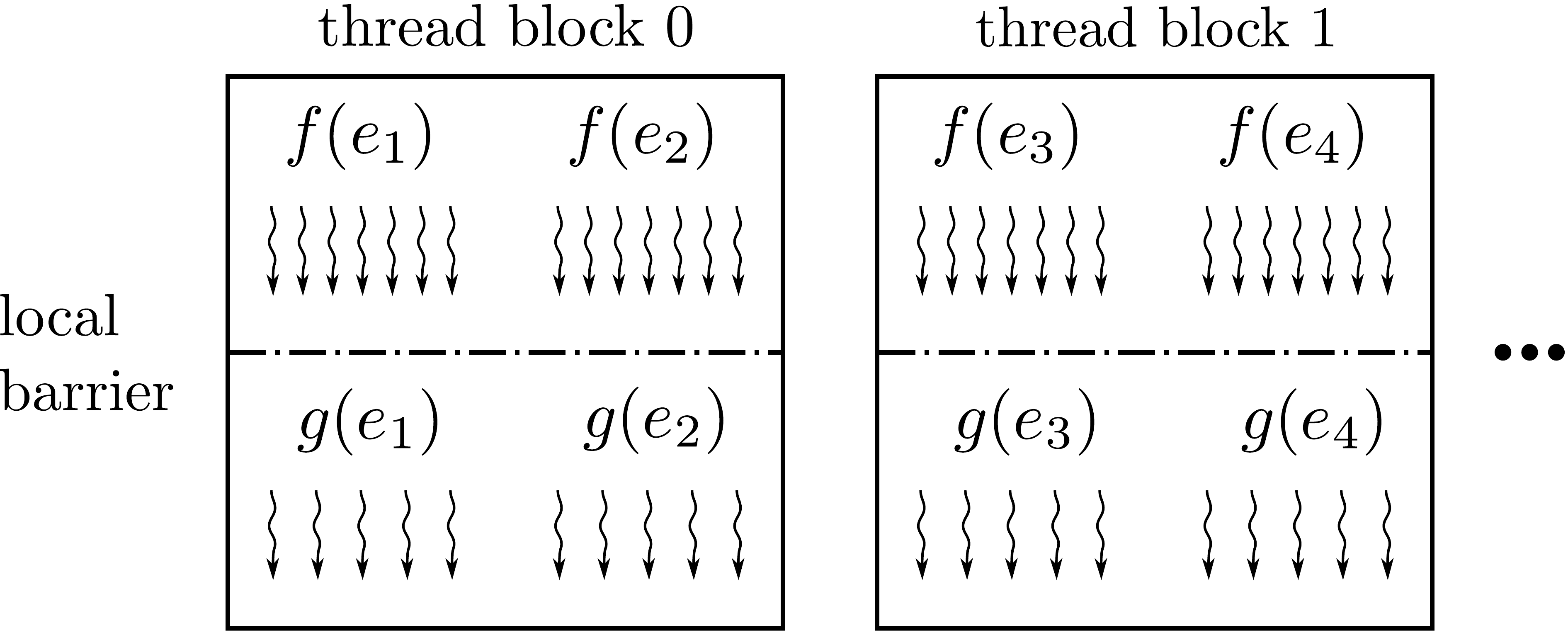}
\caption{Fused kernel performing $map(g \circ f, L)$.}
\label{fig:map-ilustration}
\end{figure}

The Figure~\ref{fig:map-ilustration} illustrates an example of kernels fusion, showing fused kernel performing $map(g \circ f, L)$. In this example, two instances $g(f(e_{2i-1}))$ and $g(f(e_{2i}))$ run in a~single thread block. As no instance is divided among thread blocks, data can be passed via on-chip memory between functions $f$ and $g$. Each function is performed by a~different number of threads, and let all threads of $f$ write a~result in this example, thus data exchanged between them cannot be placed in registers and thus must be placed in shared memory. Finally, a~local barrier has to be used.

We does not consider a fusion of functions with different nesting depth, as it yields redundant execution of functions with lower nesting depth. We note that all rules discussed in this chapter are same for nested and unnested map and reduce.

\subsection{BLAS Functions Expressed as Map and Reduce Calls}
\label{sect:blas_mapred}

In this paper, we use several BLAS functions or sequences of BLAS function calls as demonstration of described mechanisms used in our compiler. Thus, we first describe how to express BLAS functions to be usable with our compiler.

Many BLAS functions can be expressed as a~map, a~reduce or a~combination of the two. Thus, it is possible to fuse them automatically, \ie{} to perform several BLAS functions as well as their fragments within a~single kernel.

BLAS-1 implements vector-vector operations. Each vector can be expressed as a~list of elements (scalars, or small sub-vectors). BLAS-1 operations process vector elements independently (\ie{} perform map), perform reductions, or a combination of the two. For example, DOT kernel ($z \leftarrow x^T y$) multiplies each element from vector $x$ by corresponding element of vector $y$ (map) and sums the  results of multiplication over all elements (reduce). Currently, we have not implemented all BLAS-1 operations, however, we are not aware of any BLAS-1 function that cannot be implemented in the described model.

A more complicated situation arises when BLAS-2 functions, which implement matrix-vector operations, are considered. 
The matrix can be seen as a list of vectors, where vectors represent rows or columns. Then, we could implement BLAS-2 functions as a~mapping of scalar-vector functions (where scalars are elements of vectors and vectors are elements of matrices).
However, vectors representing matrix columns or rows may be too large to be stored in on-chip memory, which makes fusions impossible. To overcome this limitation, we introduce nested calls of map and reduce, which allows us to further divide vectors representing matrix rows or columns to smaller elements and process these elements with map or reduce. 

Consider the matrix-vector multiplication $y = Ax$ as an example of a~BLAS-2 function. In its computation, for the $i$-th row of $A$, the dot product of the row and $x$ is computed and stored to the $i$-th position in $y$.

Without nesting, we represent $y = Ax$ as:
\begin{equation}
\label{eq:sgemv_unnested}
y = map(dotprod(A_i, x), A)
\end{equation}
where $A$ is a~list of rows $A_1, A_2, \dotsc, A_n$,  and $x$ is a~list of elements of the vector $x$. The single instance of function $dotprod$ computes dot product of row $A_i$ and vector $x$, which may be too large for on-chip memory.

Using nested map and reduce, $y = Ax$ can be expressed as:
\begin{equation}
\label{eq:sgemv}
y = map(reduce(+, map(\cdot, A_i, x)), A)
\end{equation}
where $A$ is a~list of rows $A_1, A_2, \dotsc, A_n$, each row $A_i$ is a~list of elements in $i-th$ row and $x$ is a~list of elements of the vector $x$.

We note that we can view dimensions of multi-dimensional structures in any order -- \eg{} matrix can be viewed as a list of rows or list of columns. Thus, we can similarly express operations for transposed matrices.

The current BLAS-2 standard cannot be fully implemented using our model based on map, reduce and their nested combination. For example, we cannot handle symmetric matrices stored in some packed format in this model. To overcome this limitation, more general data structures have to be supported by our compiler in the future.

\section{The Compilation Process}
\label{sect:compilation}

Based on observations given in Section \ref{sect:kerfus}, we have developed a source-to-source compiler, which is able to optimize sequence of kernel calls by kernel fusion. In this section, we focus on the process of creating fusions and fusion code generation and briefly describe the process of fusion space exploration, which is discussed in more detail in our previous paper \cite{fousek2011automatic}.

\subsection{Compilation Stages}

Our compiler works with a special form of kernel implementation containing CUDA code implementing some higher-order function applying some first-order function on many elements---we call this special form of a kernel \textit{elementary function}. The main purpose of the compiler is to transform a~sequence of elementary function calls into the sequence of kernel calls, where single kernel can include one or more elementary functions, maximizing performance of output code. 

Recall that the input of our compiler consists of a~high-level \textit{script} and a~\textit{library} of elementary functions. Each elementary function can be present in several alternative implementations in the library with different performance characteristics. The script calls functions from the library, thus it defines the sequence of function calls and data dependencies.

The compilation process is divided into three main stages:
\begin{itemize}
        \item parsing the script and library (reading elementary functions and their metadata);
        \item generation and search of the optimization space;
        \item code generation.
\end{itemize}

The script and metadata parsing is straightforward and is not discussed here. The optimization space exploration and code generation are discussed in the following sections in more detail.

\subsection{Generation and Search the Optimization Space}
\label{sect:statespace_overview}

The input script is parsed creating data dependency graph, where vertices represent elementary function calls and edges represents data dependency between functions. Having data dependency graph build and library data parsed, the code without fusions can be generated (\ie{} each elementary function is translated to separated kernel). However, there is usually a~large number of possible codes with fusions. Thus, the optimization space is generated and searched for the code with the best expected performance. There are three main steps in the generation of the optimization space.
\begin{itemize}
        \item Generation of \textit{fusions}. We define fusion as a~fusible subgraph of data dependency graph (selection of elementary functions, which can be safely fused without influencing input program semantics). At this step, a~space of all reasonable fusions is generated.
        \item Generation of \textit{fusion implementations}. Each fusion can be implemented in many different ways, differing in (i) calling order of functions (which can affect the amount of needed on-chip memory), (ii) chosen implementations of elementary functions, (iii) block size or (iv) number of serial iterations. At this step, implementations of each fusion are generated and their performances are predicted.
        \item Generation of \textit{combination of fusion implementations}. The combination of fusion implementations is such a~selection of fusion implementations and unfused kernels, that covers all calls of elementary functions defined in input script and maximizes predicted performance. The combination of fusion implementations can be transformed to output CUDA code covering the whole computation given by the script. The generation of combinations can be repeated many times (omitting previously selected combinations) to allow empirical search for output code with the best performance.
\end{itemize}

During the generation of the optimization space, some implementations are automatically pruned, \eg{} fusions which does not spare memory transfers or  fusion implementations which use larger amount of on-chip memory per instance than another implementation of same fusion. After the pruned optimization space is generated, the performance of each fusion implementation is predicted. The fusion implementations with poor predicted performance are not definitely pruned---when sufficient number of combinations of fusion implementations is generated, they are used in some combination.

The performance prediction works as follows. 
The basic idea of our performance prediction method is to sum previously benchmarked running times of routines according to the fusion implementation for which the performance is being predicted.
More precisely, the running time of each routine is measured in simulated fusion environment -- certain ranges of the number of instances per block, sequential iterations and additionally allocated shared memory (which simulates additional data used within the fusion). When the runtime of some fusion $F$ is predicted, the runtimes of routines used in $F$, matching the fusion environment of $F$, are summed. The time of data transfers (\ie{} load and store routines) $t_t$ and computation (\ie{} compute routines) $t_c$ are summed separately and the predicted runtime is computed as $max(t_t, t_c)$. Thus, we assume full overlap of computation and data transfers. This is inaccurate when occupancy is low and the overlapping ability is reduced. However, in that case the timing of each routine is also poor, thus fusions with low occupancy should not be favored even when full overlap is expected.

Note that the benchmarking of routines is performed once per routine per GPU architecture and not at the time of compilation. The demand of benchmarking new routines is compensated for by the low sensitivity of our compiler to GPU architecture changes -- as our performance prediction is based on empirical measurements, the new GPU architecture needs only re-benchmarking of routines rather than a re-design or re-parametrization of the prediction method.

\subsection{Creating Kernels from Elementary Functions}

Recall, that our compiler is not able to fuse generic kernels implemented in C for CUDA, but works with elementary functions. 
In fact, elementary function used by our compiler contains nearly complete code of unfused kernel---however, it must fulfill several requirements described below.

The elementary function is implemented to perform some higher-order function applying some first-order function on many elements (in current implementation, higher-order function can be map, reduce or nested combination). The single instance of elementary function performs the first-order function to some input elements generating output element, \ie{} when elementary function performs $map(f, L)$, single instance performs $f(e_i)$, when elementary function performs $reduce(\oplus, L)$, single instance performs $e_{2i-1} \oplus e_{2i}$.

To be usable with our compiler, the elementary function is associated with \textit{metadata}, which defines its properties, such as parallelism requirements or implemented higher-order function. Each elementary function has to be implemented in several \textit{routines} (functions called from CUDA code):
\begin{itemize}
        \item \textit{load} (separate for each input type), loads input data stored in global memory into on-chip memory;
        \item \textit{compute} performs computation on data in on-chip memory;
        \item \textit{store} stores data from on-chip memory into global memory.
\end{itemize}

The decomposition of elementary function into routines is the core principle which significantly simplifies the code generation. The kernel is created as a~sequence of load, compute and store routine calls. When some functions are fused, the stores and loads for elements that remain in on-chip memory are not called and the remaining calls are glued into single kernel (see Figure~\ref{fig:code_fusion} for illustration of a simple fusion).

\begin {figure}[h]
  \begin{center}
    \includegraphics[width=.5\hsize]{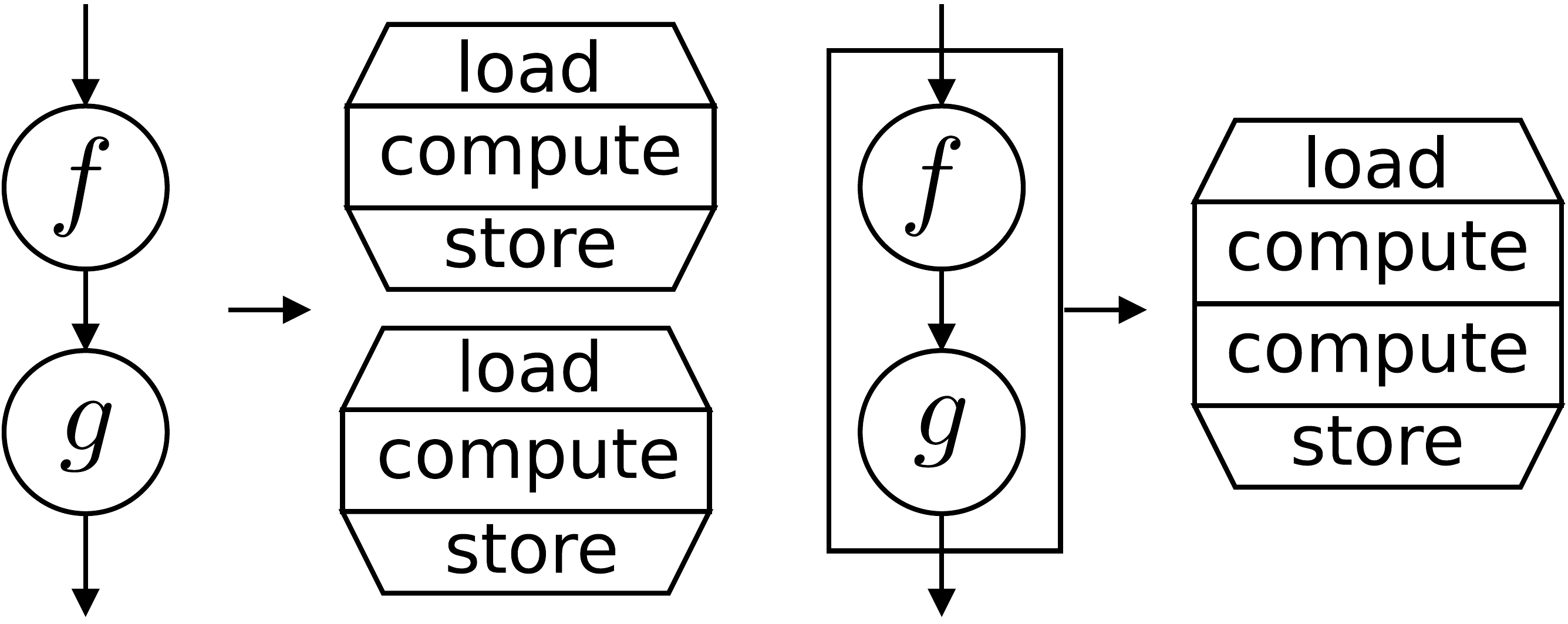}
  \end{center}
  \caption{Illustration of a simple fusion.}
  \label{fig:code_fusion}
\end {figure}

The compiler generates routine code, kernels calling these routines and a code encapsulating kernels allowing to empirically search for the most efficient one.

\subsubsection{Routine Code Generation}

First, the compiler generates routines: it copies their code from the library, assigns values to macros and modifies local memory addressing, when registers are used to store input or output elements. Macros in routines have prescribed names and are used for the thread block size and number of iterations. They are used to allow the CUDA compiler to evaluate expressions which use them at compile time or unroll small loops.

\subsubsection{Main Kernel Structure}

When the kernel code is created from elementary functions, the compiler knowis the type of the higher-order function which is implemented by the elementary function (the type of higher-order function is defined in metadata). It allows the compiler to (i) generate the computation of thread and block indices and configure the grid size, (ii) force a~global barrier before the result of the reduction is used\footnote{This is trivially fulfilled in code generation stage, as outputs of all reductions are used outside of the fusion implementation performing the reduction, thus the global barrier is performed by finishing the kernel.} and (iii) correctly place loads of invariant variables and stores of accumulable variables (\ie{} variables that can be accumulated outside of the cycle performing sequential iterations).

The unnested function runs in a~1D grid. When more than one serial iteration is performed, the grid is adequately shrunk and block indices are recomputed in each iteration, simulating the execution of the full-sized grid. For the nested functions (recall that only nesting level 2 is supported in the current implementation), a~2D grid is used, and iterations shrink the grid in one dimension, working similarly to unnested functions. In the following paragraphs, we show structure of the generated code.

\begin{algorithm}[subsection]
\caption{Schema of kernel}
\label{alg:kernel}
{
\begin{algorithmic}[1]
{
\STATE allocate variables in shared memory
\STATE create arrays in registers
\STATE compute thread and block indices
\STATE load invariant data
\STATE clear outputs of accumulated reductions
\FOR{$i = 0; i < iters; i++$}
\STATE call non-invariant load, compute and store routines
\STATE recompute block indices
\ENDFOR
\STATE call stores of accumulated reductions
}
\end{algorithmic}
}
\end{algorithm}

Algorithm~\ref{alg:kernel} sketches the basic structure of the generated (fused or unfused) kernel. All data exchanged between routines via shared memory are allocated at the beginning of the kernel (line 1). Elements in shared memory can overlap when possible to spare shared memory usage~\cite{fousek2011automatic}. This is technically realized by allocating one large array and creating pointers into this array, representing data elements. For data elements stored in registers, local arrays are defined (line 2). The size of local arrays is set to the size of one element regardless of the fraction of element used by one thread~\cite{filipovic2012automatically}.

The thread and block indices are set at line 3 according to real thread and block indices for the kernel. When some routine within the kernel needs different parallelism, indices are recomputed before this routine is called.

For nested map or reduce, some input data elements can be invariant across iterations (\eg{} for matrix-vector multiplication, a~sub-vector can multiply several matrix tiles), thus invariant loads are called before the loop (line 4). Both nested and unnested variants of reduce can accumulate their result across iterations, thus their results are cleared before the loop (line 5) and stored after the loop finishes (line 10). The rest of the routines are called within the loop (line 7) according to selected calling order and the block indices are recomputed at the end of each iteration (line 8).

Note that fusion of nested and unnested functions are not efficient, as it results in repetition of unnested operations and hence does not spare global memory transfers. Therefore, our compiler does not fuse functions with different nesting level. Consequently, compute routines are always performed within the loop, as no result of a~compute routine performed within the fusion is invariant across loop iterations.

\subsubsection{Generation of Routine Call}

\begin{algorithm}[subsection]
\caption{Schema of routine call}
\label{alg:routine}
{
\begin{algorithmic}[1]
{
\STATE call local barrier
\STATE clear output of the reduction
\IF{thread participates}
\STATE recompute thread indices
\STATE call routine
\ENDIF
}
\end{algorithmic}
}
\end{algorithm}

In Algorithm~\ref{alg:kernel}, routines are called at lines 4, 7 and 10. The more detailed schema of a~generated routine call is described in Algorithm~\ref{alg:routine}. First, the local barrier call can be generated. The local barrier before routine $r$ is generated, if one of the following conditions holds.
\begin{itemize}
	\item Routine $r$ accesses at least one input element $e$, that has been modified by routine $s$, and (i) thread-to-data mapping of access to $e$ is different for $r$ and $s$ and (ii) no local barrier is called between $r$ and $s$.
        \item Routine $r$ writes the element $e$ into shared memory, and $e$ overlaps with another element $e'$, that is accessed after last synchronization called before $r$.
\end{itemize}

The first condition ensures that all words of the element $e$ are written into shared memory before they are read by $r$, when thread-to-data mapping is different in writing and reading the element. The second condition provides synchronization of all warps before element $e'$ is rewritten by $e$ to ensure that all routines accessing $e'$ are finished before its rewriting.


When the routine performing reduction is to be called and its output is not accumulated among iterations, the code clearing its output is generated at line 2.

If the routine is performed by a~lower number of threads than it is available within the kernel, line 3 reducing the parallelism is generated. The code reducing parallelism is created to keep maximum of warps fully utilized, \ie{} when parallelism is reduced from $m$ to $n$ threads, threads $<0, \dotsc, n-1>$ continue in computation whereas threads $<n, m-1>$ stall.

The thread indices recomputation (line 4) is generated when parallelism is reduced, or when the thread arrangement of the routine differs from the thread arrangement of the fusion (\eg{} a~routine need a block of $9 \times n \times 1$ threads for $n$ instances, whereas the fused kernel uses block of $3 \times 3 \times n$ threads, \ie{} the same number of threads, but different indexing). The compiler generates the indexing computation that maps adjacent indices to adjacent threads to create at most one under-populated warp. Moreover, as it knows the number of threads in each dimension required by routines in compile time, it optimizes the number of arithmetic operations needed to recompute indices. After the parallelism is restricted and indices are recomputed, the routine can be called (line 5) with new indices.

\subsection{An Example of Code Generation}

To demonstrate the compiler's features described above, we have chosen the computation of BiCGK sequence as an example. The sequence performs
\begin{align*} 
q &= Ap \\
s &= A^Tr 
\end{align*}

It demonstrates kernel fusion ($q = Ap$ and $s = A^Tr$ are implemented as separated elementary functions) as well as working with nested operation.

Recall that the vector and matrix elements can be represented by a~single number, or some larger structure. We are using a~32-number sub-vector as a~vector element and $32\times32$ tile as a~matrix element. These element sizes allow us to write efficient elementary functions, such as $q = Ap$ or $s = A^Tr$, where single instance multiplies $32\times32$ matrix tile by sub-vector of size 32, giving good opportunity for manual optimizations. It implies that the size of A (and consequently all vectors) must be padded to 32 in each dimension.

\begin{lstlisting}[caption={Script performing BiCGK sequence},mathescape,escapeinside={(*}{*)}, label=lst:bicgk_sctipt]
TILE32x32 A;
subvector32 p, q, r, s;

input A, p, r;

q = sgemv(A, p);
s = sgemtv(A, r);

return q, s;
\end{lstlisting}

We have implemented elementary functions \texttt{sgemv} ($q = Ap$) and \texttt{sgemtv} ($s = A^Tr$). The sctipt performing BiCGK sequence is listed in Listing~\ref{lst:bicgk_sctipt}.

\begin{lstlisting}[caption={Routines performing $q = Ap$},mathescape,escapeinside={(*}{*)}, label=lst:sgemv_routines]
__device__ void d_sgemv_1_load_1(TILE32x32 A, TILE32x32 s_A, 
                  int tx, int ty, int bx, int by, int sx){
        #pragma unroll
        for (int j = 0; j < 32; j+=SGEMV_1_BY)
                s_A[ty*33+tx+j*33] = A[(by*32+ty+j)*sx*32 + bx*32+tx];
}

__device__ void d_sgemv_1_load_2(subvector32 x, subvector32 s_x, 
                  int tx, int ty, int bx, int by){
        if (ty == 0)
                s_x[tx] = x[bx*32+tx];
}

__device__ void d_sgemv_1_compute(TILE32x32 s_A, subvector32 s_x, subvector32 s_y, 
                  int tx, int ty){
        float tmp = 0.0f;
        #pragma unroll
        for (int j = 0; j < 32; j+=SGEMV_1_BY)
                tmp += s_A[tx*33+ty+j]*s_x[ty+j];
        atomicAdd(s_y+tx, tmp);
}

__device__ void d_sgemv_1_save(subvector32 s_y, subvector32 y, 
                  int tx, int ty, int bx, int by){
        if (ty == 0)
                atomicAdd(y+by*32+tx, s_y[tx]);
}
\end{lstlisting}

The CUDA code of all routines of elementary function \texttt{sgemv} is listed in Listing~\ref{lst:sgemv_routines}. There are two load routines (one for matrix tile, one for sub-vector), one store routine (saving sub-vector resulting from the reduction) and one compute routine, multiplying matrix tile with sub-vector. The macro \texttt{SGEMV\_1\_BY} is expanded to the selected $y$-size of the block and the function is implemented to run only in single instance per block (as there is enough parallelism, the execution of multiple instances per block is not necessary, contrary to unnested functions). As it can be seen, the code of \texttt{sgemv} is quite low-level, but still reasonably simple.

The metadata are associated with CUDA code of \texttt{sgemv}, determining the parallelism required by single instance of the function, higher-order function and data padding. Optionally, access pattern defining thread-to-data mapping can be defined~\cite{filipovic2012automatically}. The demand for writing these function's properties into metadata brings no significant programming overhead, as they have to be known to the programmer implementing the function.

\begin{algorithm}[h]
\caption{Fused $q = Ap, s = A^Tr$}
\label{alg:bicgk}
{
\begin{algorithmic}[1]
{
\STATE allocate $A_l, p_l, q_l, r_l, s_l$ in shared memory
\STATE compute thread indices
\STATE compute tile indices $x \leftarrow block.x, y \leftarrow i \cdot block.y$
\STATE $p_l \leftarrow load(p, x)$
\STATE $s_l \leftarrow 0$
\FOR{$y' = y; y' < min(n, y+i);$}
\STATE $r_l \leftarrow load(r, y')$
\STATE $A_l \leftarrow load(A, x, y')$
\STATE $s_l \leftarrow compute\_gemtv(A_l, r_l, x, y')$
\STATE $q_l \leftarrow 0$
\STATE $q_l \leftarrow compute\_gemv(A_l, p_l, x, y')$
\STATE $q \leftarrow store(q_l, y')$
\STATE $y' \leftarrow y'+1$
\ENDFOR
\STATE $s \leftarrow store(s_l, x)$
}
\end{algorithmic}
}
\end{algorithm}

The pseudo-code of the generated fused kernel of BiCGK sequence is listed in Algorithm~\ref{alg:bicgk}. The algorithm has several inputs: $A$ is an $m \times n$ matrix, $p, s$ are vectors of size $m$, and $q, r$ are vectors of size $n$. Load, compute and store routines, which are called in the generated code, are present in the library of elementary functions.  In the optimization space searching phase, the compiler has decided to perform several serial iterations in each instance, thus, the \texttt{for} loop going over several matrix tiles is to be generated in the kernel.

The code generation works as follows. The compiler generates a~shared memory allocation for all on-chip variables. Each variable in shared memory can be padded---in this example, $A$ is allocated as array of size $33\times32$ to allow conflict-free parallel access to columns. After memory allocation, the compiler generates the computation of the thread indices and block indices $x, y$, where $y$ is stridden according to number of serial iterations $i$. When indices are computed, the routines can be called. The local part of vector $p$ loaded into $p_l$ is invariant across iterations, and the output of the partial reduction $s_l$ can be accumulated across iterations. Thus, the compiler puts loading of $p_l$ and zeroing of $s_l$ before the loop. Within the \texttt{for} loop, the local part of $r$ and $A$ are loaded to $r_l$ and $A_l$, and $A^T_l r_l$ is computed and added to $s_l$ (line 9). To compute $A_l p_l$ (line 11), $p_l$ is zeroed (line 10) and stored (line 12) after the computation in each iteration. When all the iterations are finished, accumulated result in $s_l$ is stored. We note that for simplicity, local synchronizations are not shown in the pseudo code. 

\begin{figure}[h]
\centering
\includegraphics[width=.35\hsize]{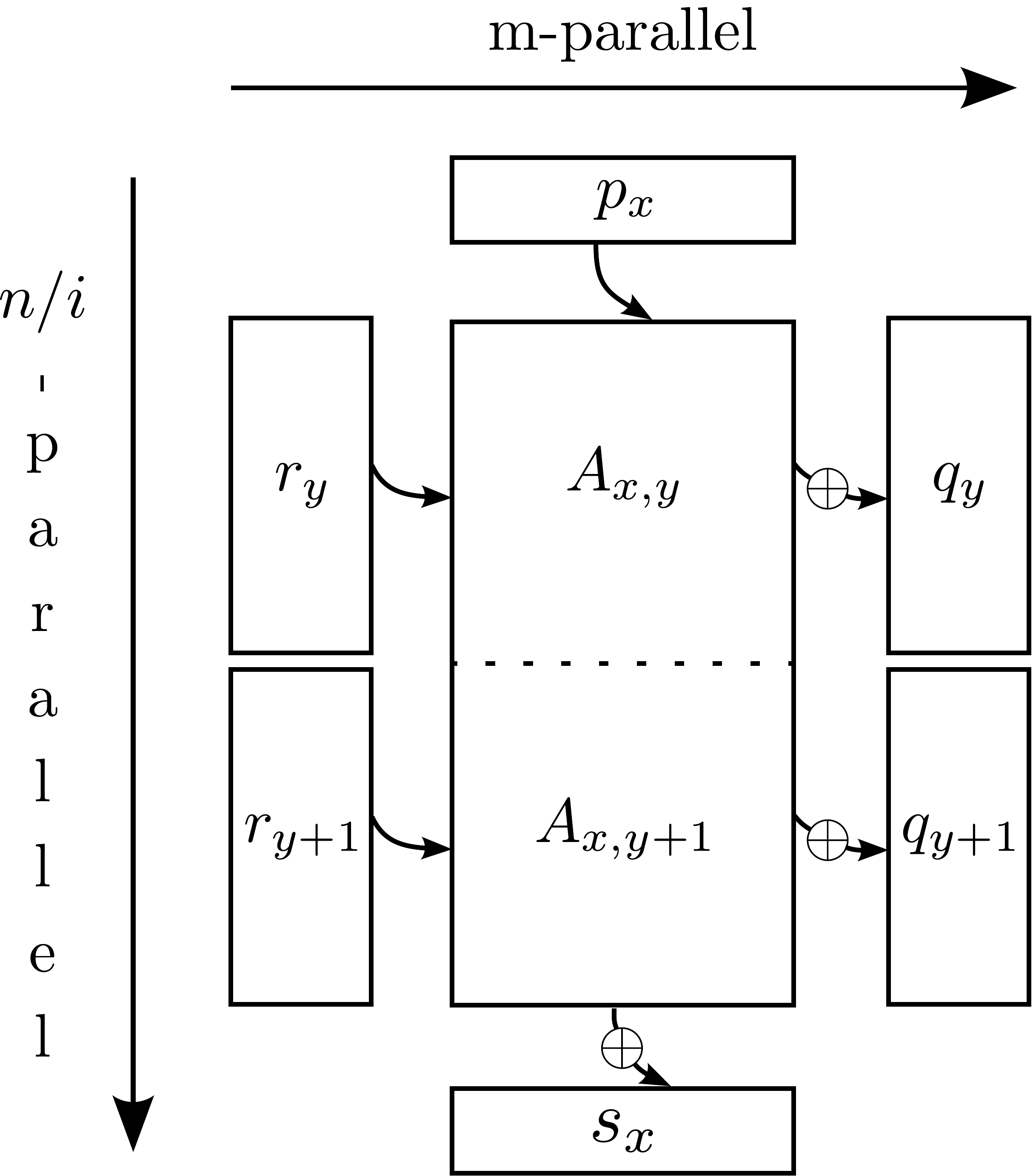}
\caption{A data usage of a~single instance of the fused BiCGK.}
\label{fig:gemv-ilustration}
\end{figure}

The data movement in the computation is illustrated in Figure~\ref{fig:gemv-ilustration}. The single instance of BiCGK processes two tiles of $A$ in depicted example ($i=2$), thus two sub-vectors of vectors $r, q$, and one sub-vector from both $p$ and $s$ are moved between global and on-chip memory. The instances are created $m \times \frac{n}{2}$ times over the matrix.

The important property of the algorithm described above is that $Ap$ can be fused together with $A^Tr$, although the dot products of the multiplied vectors and matrix $A$ are performed using rows as well as columns of the matrix---the only difference is in the placing of routines call with respect to the loop (for invariants or accumulable output).

\section{Evaluation}
\label{sect:eval}

In this section, the optimization of sequences of BLAS calls is evaluated. First, various sequences of linear algebra kernels used in our experiments are defined and the possibilities for optimizing them are analyzed. Second, the performance of implementations generated by our compiler is evaluated and compared with implementations using CUBLAS. Finally, the accuracy of the performance prediction method and compiler timing is analyzed.

\subsection{Experiment setup}

To test the code efficiency of our compiler, we have used the same sequences as in \cite{belter2009automating}, which are specified in Table~\ref{tab:kernels}. These sequences are a representative selection of generally interesting operations, where many of them have important applications (BiCGK is used in biconjugate gradient method, ATAX in normal equations computation), are added to the updated BLAS specification (AXPYDOT, SGEMVT, GEMVER, GESUMMV, WAXPBY) \cite{blackford2002updated}, are in original BLAS specification (SGEMV and SSCAL) or are generally usable (MADD, VADD). Some of these sequences can be significantly improved by fusions whereas others cannot. The adoption of sequences from~\cite{belter2009automating} allows us to compare effect of fusion on two different processors---multi-core CPU and many-core GPU. The only difference between our sequences and those used in \cite{belter2009automating} is in the floating point accuracy---we have used the single precision version of all sequences, whereas in~\cite{belter2009automating} double precision has been used\footnote{Note that the selection of different precision should not affect comparison of speedups reached by our compiler and BTO BLAS. Although double amount of data are transfered when double precision is used, the CPU SSE peak performance in double precision is a~half of single precision performance, thus the ratio of memory to arithmetic throughput is same for both single and double precisions and therefore the effect of fusions should be same.}. 

\begin{table}
	\centering
	\small
	\begin{tabular}{|l|l|l|}
		\hline
		Sequence & Operation & Tag \\
		\hline
		AXPYDOT & $z \leftarrow w - \alpha v$ & FS \\ 
		& $r \leftarrow z^T u$ & \\
		ATAX & $ y \leftarrow A^T A x$ & \\
		BiCGK & $q \leftarrow A p$ & F \\
		& $s \leftarrow A^T r$ & \\
		SGEMV & $z \leftarrow \alpha A x + \beta y$ & B \\ 
		SGEMVT & $x \leftarrow \beta A^T y + z$ & (S) \\
		& $w \leftarrow \alpha A x$ & \\
		SSCAL & $x \leftarrow \alpha x$ & B \\
		GEMVER & $B \leftarrow A + u_1 v_1^T + u_2 v_2^T$ & FS \\
		& $x \leftarrow \beta B^T y + z$ & \\
		& $w \leftarrow \alpha B x$ & \\
		GESUMMV & $y \leftarrow \alpha A x + \beta B x$ & (F) \\
		MADD & $C \leftarrow A + B$ & S \\
		VADD & $x \leftarrow w + y + z$ & FS \\
		WAXPBY & $w \leftarrow \alpha x + \beta y$ & F \\
		\hline
	\end{tabular}
	\caption{Sequences used in our performance study, adopted from \cite{belter2009automating}. Tags: F=improvable by the fusion, S=improvable by kernel specialization, B=equivalent of CUBLAS kernel.}
	\label{tab:kernels}
\end{table}

We have assigned tags to each sequence in Table~\ref{tab:kernels}. These tags indicate optimizations that our compiler is able to perform. Tag F indicates that fusion can be used to improve performance, tag S indicates that more specialized kernels that save some work compared to CUBLAS can be generated. Finally, tag B indicates sequences that have their equivalents in CUBLAS, thus any optimization that can be used by our compiler can also be implemented in CUBLAS. When some tag is enclosed in brackets, its significance is low, \ie{} is related to BLAS-1 operations in sequences where much more time-consuming BLAS-2 operations are executed.

For some sequences, the tag assignment does not have to be straightforward, thus we discuss it in more detail.
\begin{itemize}
	\item ATAX and SGEMVT cannot be improved by fusion. In both cases, matrix A is used twice, but a global barrier is needed between uses of A, and thus must be used in separate kernels.
	\item GESUMMV can spare the reading of vector $x$ when it is performed in a single kernel. However, because of reading the matrices $A$ and $B$, the amount of data transfer is almost the same in the fused and unfused versions.
	\item All sequences with the S tag require memory copy or cleaning in the CUBLAS implementation because of the in-place implementation of some CUBLAS kernels, whereas kernels generated by our compiler can work out-of-place.
\end{itemize}

\subsection{Performance Results}

All experiments have been performed on a machine equipped with an Intel Core2 Q9550 (2.83\,GHz), 8\,GB RAM, and a GeForce GTX 480. Ubuntu 10.04 with CUDA Runtime 5.0 and Driver 304.54 have been installed.

Table~\ref{tab:perf} compares performance of code generated by our compiler with CUBLAS implementations.

\begin{table}
        \centering
	\small
        \begin{tabular}{|l|l|l|l|l|}
                \hline
                Sequence & Our compiler & CUBLAS & Speedup & Tag \\
		\hline
                AXPYDOT & 38.3\,GFlops & 19.7\,GFlops & $1.94\times$ & FS \\
                ATAX & 73.5\,GFlops & 71.5\,GFlops & $1.03\times$ &  \\
                BiCGK & 115\,GFlops & 71.5\,GFlops & $1.61\times$ & F \\
                SGEMV & 73.3\,GFlops & 69.9\,GFlops & $1.05\times$ & B \\
                SGEMVT & 73.3\,GFlops & 71.5\,GFlops & $1.03\times$ & (S) \\
                SSCAL & 18.2\,GFlops & 17.3\,GFlops & $1.05\times$ & B \\
                GEMVER & 83.4\,GFlops & 31.9\,GFlops & $2.61\times$ & FS \\
                GESUMMV & 73.4\,GFlops & 73.1\,GFlops & $1\times$ & (F) \\
                MADD & 11.3\,GFlops & 7.68\,GFlops & $1.47\times$ & S \\
                VADD & 20.0\,GFlops & 8.84\,GFlops & $2.26\times$ & FS \\
                WAXPBY & 36.4\,GFlops & 18.9\,GFlops & $1.93\times$ & F\\
                \hline
        \end{tabular}
        \caption{Performance comparison of generated and CUBLAS implementations of studied sequences.}
        \label{tab:perf}
\end{table}

In all cases, the generated implementation performs better or similarly compared to CUBLAS. Significant speedup is obtained in the case of sequences where the fusion can improve memory locality (tag F) as well as when kernel specialization is possible (tag S). Those sequences demonstrate the strength of our compiler, as they are improved by compiler's optimizations which cannot be implemented in CUBLAS (without modification of its API).

To the best of our knowledge, there is no other system allowing  fusion of BLAS functions for GPUs that could be compared with our results. Nevertheless, we can compare the relative speedup of our generated codes with relative speedup of CPU code generated by BTO BLAS~\cite{belter2009automating}, see Table~\ref{tab:fublas_comp}.

Our speedup is generally better comparing to the speedup of BTO BLAS when fusion can be used (sequences AXPYDOT, BiCGK, GEMVER, VADD, WAXPBY). Our compiler is more successful with sequences equivalent to BLAS functions (SGEMV) or sequences with reduced opportunity to be improved by fusion (GESUMMV)---in our case, the performance is comparable, whereas BTO BLAS generates slower codes. The main reason is probably that our compiler fuses hand-written routines, that can be adequately tuned. The exactly same speedup is shown in the case of MADD, where only kernel specialization takes place.

On the other hand, BTO BLAS has a~wider opportunity to enhance code performance using fusions. When the function $f$ performs reduction on each row of the matrix and the reduction's result is an input of function $g$ processing the same row, CPU is able to hold the row in the cache and reuse it after reduction finish (thus outer loops in $f$ and $g$ going over rows are fused, whereas inner loops are unfused). Considering GPU, the row needs to be partitioned among more thread blocks when is read into on-chip memory by $f$, thus thread blocks needs to be synchronized before the result of reduction is available. Our compiler performs the synchronization by the new kernel invocation, thus all on-chip data are lost before the result of the reduction is available for $g$ so no row data can be reused. The only option how to reuse row data on GPU is to use persistent threads~\cite{gupta2012study}, but it is not clear if it could have a~positive performance impact, as the inter-block synchronization is possible, but decreases the performance. The wider fusion opportunity of BTO BLAS caused better speedup in sequences ATAX and SGEMVT.

\begin{table}
        \centering
        \small
        \begin{tabular}{|l|l|l|l|}
                \hline
                Sequence & Our     & BTO BLAS & Our memory\\
                         & speedup & speedup  & bandwidth\\
                \hline
                AXPYDOT & $\mathbf{1.94\times}$ & $1.58\times$ & 153.2\,GB/s\\
                ATAX & $1.03\times$ & $\mathbf{1.37\times}$ & 147\,GB/s\\
                BiCGK & $\mathbf{1.61\times}$ & $1.5\times$ & 115\,GB/s\\
                SGEMV & $\mathbf{1.05\times}$ & $0.83\times$ & 146.6\,GB/s\\
                SGEMVT & $1.03\times$ & $\mathbf{1.29\times}$ & 146.6\,GB/s\\
                SSCAL & $1.05\times$ & n/a & 145.6\,GB/s\\
                GEMVER & $\mathbf{2.61\times}$ & $2.37\times$ & 143\,GB/s\\
                GESUMMV & $\mathbf{1\times}$ & $0.93\times$ & 146.8\,GB/s\\
                MADD & $1.47\times$ & $1.47\times$ & 135.6\,GB/s\\
                VADD & $\mathbf{2.26\times}$ & $1.83\times$ & 160\,GB/s\\
                WAXPBY & $\mathbf{1.93\times}$ & $1.88\times$ & 145.6\,GB/s\\
                \hline
        \end{tabular}
        \caption{Comparison of the speedup of sequences generated by our compiler and best cases generated by BTO BLAS and the memory bandwidth of our kernels.}
        \label{tab:fublas_comp}
\end{table}

As sequences analyzed in this section are memory-bounded even after fusion, their arithmetic throughput is far from the peak throughput of GeForce GTX 480. To determine generated kernels efficiency, we have measured their bandwidth (shown in last column of Table~\ref{tab:fublas_comp}). Note that the bandwidth of fused kernels is measured (\ie{} only bandwidth of data that are really transfered from or to global memory), which gives us the information about real efficiency of kernel implementations hiding improvements of the fusion. The maximal theoretical bandwidth of GeForce GTX480 is 177.4\,GB/s. However, this bandwidth is unreachable in practice. The most of our kernels reaches more than 75\,\% of theoretical peak, which can be considered as a very good result (there is no significant chance to improve their performance by tuning the compiler). The BiCGK kernel reaches about 65\,\% of the peak bandwidth. We have experimented manually with tuning this kernel, reaching 78\,\% of the peak bandwidth when loops of load and compute routines iterating over matrix tile are manually fused, showing further challenges in automatic kernel fusion.

\subsection{Performance Prediction Accuracy}

We have analyzed the accuracy of the performance prediction. As numerous possible implementations can be generated, good performance prediction allows the reduction of the empirical search to several promising candidates or eliminate empirical searching entirely. 

\begin{table}
        \centering
        \small
        \begin{tabular}{|l|l|l|l|l|}
                \hline
                Sequence & Impl. & Best impl. & First impl. & Worst impl. \\
                name     & count & found      & performance & performance \\
                \hline
                AXPYDOT  & 25    & 4th         & 75.2\,\%    & 34.9\,\% \\
                ATAX     & 1     & 1st         & 100\,\%     & n/a \\
                BiCGK    & 5     & 1st         & 100\,\%     & 64.0\,\% \\
                SGEMV    & 83    & 14th        & 99.2\,\%    & 97.8\,\% \\
                SGEMVT   & 41    & 5th         & 99.8\,\%     & 99.4\,\% \\
                SSCAL    & 1     & 1st         & 100\,\%     & n/a \\
                GEMVER   & 1271  & 54th        & 98.7\,\%    & 43.1\,\% \\
                GESUMMV  & 415   & 51st        & 99.6\,\%    & 94.4\,\% \\
                MADD     & 1     & 1st         & 100\,\%     & n/a \\
                VADD     & 41    & 14th        & 94.6\,\%    & 50.4\,\% \\
                WAXPBY   & 83    & 1st         & 100\,\%    & 29.3\,\% \\
                \hline
        \end{tabular}
        \caption{For each studied sequence, the count of all implementations is shown in the second column, the rank of the best generated implementation is shown in the third column, the performance of the first generated implementation relative to the best one is in the fourth column and the performance of the worst implementation relative to the best performing implementation is shown in the fifth column. To eliminate measurement error, all implementations for which performance does not differ more than 0.1\,\% are considered to have same performance.}
        \label{tab:prediction}
\end{table}

Table~\ref{tab:prediction} shows the number of possible implementations of each sequence and the rank of the fastest implementation. As we can see, the best implementation is not generated as the first one in six cases. However, the performance of the first generated implementation is reasonably close to the best one except the AXPYDOT routine (see fourth column of the table).

The selection of inefficient implementation of AXPYDOT is caused by a~systematic error in the performance prediction method which underestimates performance of fused kernels. The error is probably caused by ignoring kernel startup overhead and serial code optimizations. As performances of the fused and unfused versions of AXPYDOT are relatively close, the compiler wrongly expects unfused version to run faster.

The last column of Table~\ref{tab:prediction} shows the performance of the worst generated implementation compared to the best one. As we can see, the worst generated implementations often perform poorly, thus the sorting of possible implementations by predicted performance is crucial.



\subsection{Compilation Time}

\begin{table}
        \centering
        \small
        \begin{tabular}{|l|l|l|r|}
                \hline
                Sequence & First           & All             & Empirical  \\
                         & implementation  & implementations & search     \\
                \hline
                AXPYDOT  & 0.144\,s & 0.241\,s & 1\,m\,59\,s\\
                ATAX     & 0.137\,s & 0.144\,s & 5\,s\\
                BiCGK    & 0.140\,s & 0.164\,s & 18\,s\\
                SGEMV    & 0.152\,s & 0.900\,s & 8\,m\,22\,s\\
                SGEMVT   & 0.123\,s & 0.393\,s & 4m\,\,42\,s\\
                SSCAL    & 0.139\,s & 0.113\,s & 3\,s\\
                GEMVER   & 0.133\,s & 42.165\,s & 3\,h\,24\,m\,36\,s\\
                GESUMMV  & 0.123\,s & 5.707\,s & 48\,m\,23\,s\\
                MADD     & 0.128\,s & 0.116\,s & 4\,s\\
                VADD     & 0.133\,s & 0.248\,s & 3\,m\,3\,s\\
                WAXPBY   & 0.156\,s & 0.731\,s & 7\,m\,14\,s\\
                \hline
        \end{tabular}
        \caption{Time of compilation and empirical search for tested sequences.}
        \label{tab:fublas_comp_timing}
\end{table}

The compilation time and empirical search time are given in Table~\ref{tab:fublas_comp_timing}. As we can see, compilation time is rather same when only implementations with best predicted performance are generated. When all possible implementations (given by combinations of fusion implementations) are generated, the compilation time is still feasible: it reaches at most tens of seconds in compilation of GEMVER sequence (1271 implementations generated), less then 6 seconds in GESUMMV and less than 1 second all other sequences. However, the time for empirical search for best performing implementation increases proportionally with the number of implementations. Fortunately, the empirical search has small impact on performance and if it is used, only a~few implementations needs to be generated and benchmarked to have a good chance to find the best performing one (see Table~\ref{tab:prediction}).

\section{Conclusions and Future Work}
\label{sect:conclusion}

In this paper, we have significantly extended our approach to automatic kernel fusion by introducing fusion of reduce and nested map and reduce kernels. We have shown that kernel fusion can improve the performance of memory-bound kernels. Although the fusion of general kernels is difficult to automate, we have argued that the automation is possible and demonstrated the automation for (possibly nested) map and reduce kernels using our source-to-source compiler. The application of our compiler has been demonstrated by fusing sequences of BLAS calls, where a significant speedup comparing to CUBLAS has been observed.

We plan to focus on generalization of the presented method, allowing to fuse more types of kernels, work with irregular data structures or target multiple GPUs.
\begin{itemize}
	\item \textit{Support for more types of fusible kernels.} A more general model of temporal locality in GPU memory hierarchy could be formulated, which would allow us to handle more types of kernels, such as stencils, scatters or gathers. Supporting more general kernels significantly extends the applicability of our compiler, \eg{} in image processing, ODE solvers, or Finite Difference Method.
	\item \textit{Support for irregular data types.} The operations working with irregular data types, such as triangular or diagonal matrices, or sparse arrays, need more general higher-order functions, than are currently supported. Irregular or sparse structures enrich application area of our compiler, \eg{} for large simulation of physical phenomena.
	\item \textit{Support for multi-GPU computations.} To allow scaling of GPU applications, the workload needs to be distributed among multiple GPUs. While the distribution of map and reduce is quite straightforward, more complicated functions, such as nested map and reduce or stencils yield significantly more difficult data exchange pattern.
\end{itemize}

Besides improving the compiler, we are going to develop libraries of elementary functions. We plan to implement more functions from the BLAS standard which are fusible by the compiler and a library of linear algebra operations on small elements usable for element subroutines in FEM.

\section*{Acknowledgements}
This work was supported by Ministry of Education, Youth and Sport of the Czech Republic under the project ``CERIT Scientific Cloud'' (no. ED3.2.00/08.0144) and by Czech Science Foundation research project ``Mathematical and Engineering Approaches to Developing Reliable and Secure Concurrent and Distributed Computer Systems'' (no. GD102/09/H042). 

\bibliographystyle{siam}
\bibliography{fila}

\end{document}